\documentclass[fleqn,twoside]{article}
\usepackage{espcrc2}

\newcommand{\AmS}{{\protect\the\textfont2
  A\kern-.1667em\lower.5ex\hbox{M}\kern-.125emS}}

\title{The technique of inverse Mellin transform for processes
occurring in a background magnetic field}

\author{Guey-Lin Lin\address{Institute of Physics, National Chiao-Tung University \\
        Hsinchu 300, Taiwan}%
        \thanks{This work is supported in part by the National Science Council of Taiwan
under the grant number NSC90-2112-M009-023.}}

\begin{document}

\begin{abstract}
We develop the technique of inverse Mellin transform for processes
occurring in a background magnetic field. We show by analyticity
that the energy (momentum) derivatives of a field theory amplitude
at the zero energy (momentum) is equal to the Mellin transform of
the absorptive part of the amplitude. By inverting the transform,
the absorptive part of the amplitude can be easily calculated. We
apply this technique to calculate the photon polarization function
in a background magnetic field. \vspace{1pc}
\end{abstract}

\maketitle

\section{INTRODUCTION}

The analytic properties of scattering amplitudes in quantum field
theories are well known. They are described by the so called
cutting rules\cite{cutkosky,veltman}. With the cutting rules, the
absorptive part of a scattering amplitude can be calculated, while
the dispersive part of the amplitude is obtained from the former
by the Kramers-Kronig relation. The above procedures can be easily
implemented for any quantum field theory with a trivial vacuum.
For such a vacuum, the energy-momentum relation of an asymptotic
state is simply $E^2={\bf p}^2+m^2$ with $m$ the mass of the
asymptotic state. On the other hand, for processes occurring in a
background magnetic field, the energy-momentum relations of
asymptotic states receive corrections from the magnetic-field
effects. For instance, the energy-momentum relation of an electron
(positron) in the background magnetic field is given by
\begin{equation}
E^2_{n,s_z}=m_e^2+p_z^2+eB\left(2n+1+2s_z\right),
\label{quanta}
\end{equation}
where the background magnetic field is taken to be along the $+z$
direction, $s_z$ is the electron spin projection, and $n$ labels
the Landau levels. Due to the above energy quantization caused by
a background magnetic field, one expects the absorptive part of a
physical amplitude contains an infinite number of contributions
distinguished by their corresponding Landau levels. Hence the
calculation of the absorptive part is becoming rather involved,
and certainly the calculation of the dispersive part is even more
intricate.

In this talk, we present a new approach for computing the
absorptive part of a physical amplitude in the background magnetic
field. The idea is based upon the analyticity. For illustration,
we take the one-loop photon polarization function as an example,
with a background magnetic field along the $+z$ direction. Since
the energy of the internal electron in the photon polarization
function is given by Eq. (\ref{quanta}), the value of the photon
longitudinal momentum $q_{\parallel}^2\equiv q_0^2-q_z^2$
determines the threshold for the absorptive part. Therefore the
photon polarization function is an analytic function of the photon
longitudinal momentum $q_{\parallel}^2$ except on the positive
real axis. To obtain the photon polarization function for an
arbitrary $q_{\parallel}^2$, it suffices to know the function's
power-series expansion in $q_{\parallel}^2$ at
$q_{\parallel}^2=0$, because the analytic continuation can map the
function from the neighborhood of $q_{\parallel}^2=0$ to any value
of $q_{\parallel}^2$. A powerful way to perform this analytic
continuation is through the inverse Mellin
transform\cite{KLTa,KLTb}. With the inverse Mellin transform, one
calculates the absorptive part of the photon polarization function
with the knowledge of the above-mentioned power-series expansion.
We note that it is straightforward to obtain such a power series
because the neighborhood of $q_{\parallel}^2=0$ is free of
resonant singularities caused by the creation of electron-positron
pairs. Knowing the absorptive part, the dispersive part of the
photon polarization at any $q_{\parallel}^2$ is calculable by the
Kramers-Kronig relation.

We organize this presentation as follows: In Section 2, we
illustrate the technique of inverse Mellin transform using the
vacuum QED as an example. In Section 3, we  apply this technique
to the photon polarization function in a background magnetic
field. A short conclusion is given in Section 4.

\section{VACUUM QED}
The vacuum polarization tensor in QED is written as
\begin{equation}
i\Pi^{\mu\nu}(q)=(q^2 g^{\mu\nu}-q^{\mu}q^{\nu})i\Pi(q^2).
\end{equation}
We are interested in calculating the one-loop finite part
$\bar{\Pi}(q^2) =\Pi(q^2)-\Pi(0)$. We observe that
$\bar{\Pi}(q^2)$ satisfies the sum rule \cite{KLTa}:
\begin{eqnarray}
&&\frac{1}{n!}\left(\frac{d^n}{d(q^2)^n}
\bar{\Pi}(q^2)\right)_{q^2=0}\nonumber \\
&=& \frac{1}{\pi}\int_{M^2}^{\infty}dq^2 {\rm Im}\bar{\Pi}(q^2)
(q^2)^{-(n+1)}, \label{sum_rule}
\end{eqnarray}
where $M^2$ is the threshold energy for the absorptive part ${\rm
Im}\bar{\Pi}(q^2)$. Although the value for $M^2$ is not specified
at this moment, the procedure of analytic continuation will
automatically generate it. Let us rewrite the sum rule in
dimensionless variables:
\begin{equation}
\frac{1}{n!}\left(\frac{d^n}{dt^n}\bar{\Pi}(t)\right)_{t=0}=
\frac{1}{\pi}\int_{0}^{1}du {\rm Im}\bar{\Pi}(u)u^{n-1},
\end{equation}
where $t=q^2/M^2$ and $u=1/t$. It is easily seen that the
derivatives of $\bar{\Pi}$ at $t=0$ is proportional to the Mellin
transform of ${\rm Im}\bar{\Pi}$. Hence ${\rm Im}\bar{\Pi}$ at any
value of $t$ is obtainable by the inverse Mellin transform:
\begin{equation}
{\rm Im}\bar{\Pi}(u)=\frac{1}{2i}\int_{C}ds\, a(s) u^{-s},
\label{inverse_mellin}
\end{equation}
where $a(s)$ is the analytic continuation of $a(n)$ which appears
in the power series expansion
\begin{equation}
\bar{\Pi}(t)=\sum_{n=0}^{\infty}a(n)t^n.
\end{equation}
For the current case, $a(n)$ is given by
\begin{equation}
a(n)=-\frac{2\alpha}{\pi}\frac{\Gamma^2(n+2)}{n\Gamma(2n+4)}
\left(\frac{M^2}{m_e^2}\right)^n.
\end{equation}
Applying Eq. (\ref{inverse_mellin}), we obtain
\begin{eqnarray}
{\rm
Im}\bar{\Pi}(u)=&-&\frac{\alpha}{3}\sqrt{1-\frac{4m_e^2}{M^2}u}
\left(1+\frac{2m_e^2}{M^2}u\right)\nonumber \\
&\times&\Theta\left(1-\frac{4m_e^2}{M^2}u\right).
\end{eqnarray}
We like to remark that, with $u=1/t=M^2/q^2$, ${\rm
Im}\bar{\Pi}(u)$ is independent of the undetermined threshold
scale $M^2$, and it agrees with the known result obtained by
applying the cutting rules. Having obtained ${\rm
Im}\bar{\Pi}(u)$, one can calculate ${\rm Re}\bar{\Pi}(u)$ by the
Kramers-Kronig relation.

\section{QED IN A BACKGROUND MAGNETIC FIELD}
The photon polarization function in a background magnetic field is given by the following
proper time representation\cite{schwinger,tsai}:
\begin{eqnarray}
&&\Pi_{\mu\nu}(q)=-{e^3B\over (4\pi)^2}\int_0^{\infty}ds
\int_{-1}^{+1} dv \times \nonumber \\
&&\{ e^{-is\phi_0}[T_{\mu\nu}N_0
-T_{\parallel,\mu\nu}N_{\parallel}
+T_{\bot,\mu\nu}N_{\bot}]\nonumber \\
&&-e^{-ism_e^2}(1-v^2)T_{\mu\nu}\}, \label{proper_t}
\end{eqnarray}
where the photon momentum has been decomposed into
 $q_{\parallel}^{\mu}= (\omega, 0, 0, q_z)$ and
$q_{\bot}^{\mu}= (0, q_x, q_y, 0)$; while
$T_{\mu\nu}=(q^2g_{\mu\nu}-q_{\mu}q_{\nu})$,
$T_{\parallel,\mu\nu}=(q_{\parallel}^2g_{\parallel\mu\nu}-
q_{\parallel\mu}q_{\parallel\nu})$, and
$T_{\bot,\mu\nu}=(q_{\bot}^2g_{\bot\mu\nu}-
q_{\bot\mu}q_{\bot\nu})$. The phase $\Phi_0$ and the functions
$N_0$, $N_{\parallel}$ and $N_{\bot}$ are given by
\begin{equation}
\phi_0=m_e^2-{1-v^2\over 4}q_{\parallel}^2-{\cos(zv)-\cos(z)\over 2z\sin(z)}
q_{\bot}^2
\end{equation}
with $z=eBs$, and
\begin{eqnarray}
N_0&=&{\cos(zv)-v\cot(z)\sin(zv)\over \sin(z)},\nonumber \\
N_{\parallel}&=&-\cot(z)\left(1-v^2+{v\sin(zv)\over
\sin(z)}\right)+\nonumber \\
&&{\cos(zv) \over \sin(z)},\nonumber \\
N_{\bot}&=&-{\cos(zv)\over \sin(z)} +{v\cot(z)\sin(zv)\over
\sin(z)}+\nonumber \\
&&{2\left(\cos(zv)-\cos(z)\right)\over \sin^3(z)}.
\end{eqnarray}
The two independent eigenmodes of the above polarization tensor
are $\epsilon^{\mu}_{\parallel}$ and $\epsilon^{\mu}_{\bot}$ which
are respectively parallel and perpendicular to the plane spanned
by the photon momentum ${\bf q}$ and the magnetic field ${\bf B}$.
They obey the dispersion equations $q^2+\Pi_{\parallel}=0$ and
$q^2+\Pi_{\bot}=0$ respectively with
$\Pi_{\parallel,\bot}=\epsilon^{\mu}_{\parallel,\bot}
\Pi_{\mu\nu}\epsilon^{\nu}_{\parallel,\bot}$. It turns out that
$\Pi_{\parallel}$ and $\Pi_{\bot}$ are proportional to
$N_{\parallel}$ and $N_{\bot}$ respectively. We shall not discuss the
contribution by $N_0$ since it does not correspond to an independent eigenmode.

For simplicity, we shall only focus on the calculation of
$\Pi_{\parallel}$. In particular, we only illustrate the procedure
of calculating the partial contribution $\Pi_{\parallel}^A$ given
by
\begin{eqnarray}
\Pi_{\parallel}^A=&-&\frac{\alpha\omega^2\sin^2\theta}
{4\pi}\int_0^{\infty}dz\int_{-1}^{+1}dv\nonumber \\
&\times&\exp(-is\phi_0) \frac{\cos (zv)}{\sin (z)},
\end{eqnarray}
where the integrand $\cos (zv) / \sin (z)$ is taken from the
second term of $N_{\parallel}$, and $\theta$ is the angle between
${\bf q}$ and ${\bf B}$.

The scalar function $\Pi_{\parallel}^A$ is easy to calculate only
for $q_{\parallel}^2< 4m_e^2$, since, in this momentum region, the
contour rotation $s\to -is$ is allowed\cite{tsai}. The oscillating
trigonometric functions are then rotated into hyperbolic
functions. Similar to the previous section, one writes
$\Pi_{\parallel}^A$ as a power series in $q_{\parallel}^2$:
\begin{equation}
\Pi_{\parallel}^A=-\frac{\alpha\omega^2\sin^2\theta}{4\pi}
\sum_{n=0}^{\infty}a(n,q_{\bot}^2)
\left(\frac{q_{\parallel}^2}{4m_e^2} \right)^n.
\end{equation}
Following Eq. (\ref{inverse_mellin}), we calculate ${\rm
Im}\Pi_{\parallel}^A$ by
\begin{equation}
{\rm
Im}\Pi_{\parallel}^A(v)=\frac{i\alpha\omega^2\sin^2\theta}{8\pi}
\int_{C}ds\,
a(s,q_{\bot}^2) v^{-s},
\end{equation}
with $a(s,q_{\bot}^2)$ the analytic continuation of
$a(n,q_{\bot}^2)$, and $v=4m_e^2/q_{\parallel}^2$. We arrive at
\begin{eqnarray}
{\rm Im}\Pi_{\parallel}^A &=& \frac{2\alpha
eB\omega^2\sin^2\theta}{q_{\parallel}^2}\sum_{l_1=1,l_2=0}^{\infty}
T_{l_1,l_2}(q_{\bot}^2)\nonumber \\
&\times&\frac{\Theta(1-\frac{4\lambda
eB}{q_{\parallel}^2}+\frac{4\rho^2
e^2B^2}{q_{\parallel}^4})}{\sqrt{1-\frac{4\lambda
eB}{q_{\parallel}^2}+\frac{4\rho^2 e^2B^2}{q_{\parallel}^4}}},
\label{root}
\end{eqnarray}
where $\lambda=l_1+l_2+m_e^2/eB$, $\rho=l_1-l_2$, and the step
function indicates the threshold for creating an electron-positron
pair occupying the $l_1$-th and $l_2$-th Landau levels. The
detailed form of $T_{l_1,l_2}(q_{\bot}^2)$ is given in
Ref.~\cite{KLTb}. With ${\rm Im}\Pi_{\parallel}^A$ determined,
${\rm Re}\Pi_{\parallel}^A$ can once more be calculated by the
Kramers-Kronig relation.
\section{CONCLUSION}
In this talk, we have demonstrated the usefulness of analyticity
for computing field theory amplitudes. We observed that a field
theory amplitude in the neighborhood of zero energy (momentum)
determines the amplitude at the arbitrary energy (momentum). The
latter can be calculated from the former by the inverse Mellin
transform and the Kramers-Kronig relation. We also showed that
such an approach is very appropriate for a process occurring in a
background magnetic field, where an infinite number of thresholds
occur in the amplitude of the process.

\end{document}